\documentstyle[12pt,a4,epsfig,amsbsy,amssymb,cite]{article}

\def \logo {\vbox to 10.5 mm{
\hbox {}
\hbox to 59 true  mm
{
\def\magps { }
\ifnum\mag=1000 \def\magps {1.2 1.2
scale } \fi
\hfil}\vfil}}

\newcommand{\ee}{$\mbox{e}^+\mbox{e}^-$}
\newcommand{\WW}{$\mbox{W}^+\mbox{W}^-$}
\newcommand{\qqqq}{$q \bar {q} q \bar {q}$}
\newcommand{\qql}{$q \bar {q} l \nu$}

\newcommand{\x}{$\xi$}
\newcommand{\BE}{Bose-Einstein }
\newcommand{\beq}{\begin{equation} }
\newcommand{\eeq}{\end{equation} }
\newcommand{\bea}{\begin{eqnarray} }
\newcommand{\eea}{\end{eqnarray} }
\begin{document}

\begin{titlepage}

\logo

\vspace*{1.5cm}

\begin{flushright}
MAN/HEP/2001/2
\\ April 30, 2001
\end{flushright}

\vspace*{1.5cm}

\begin{center}
\Large{\bf Event weights for simulating
Bose-Einstein correlations}
\end{center}

\normalsize
\begin{center}
{\bf
V.~Kartvelishvili}{\footnote{Present address: Department of Physics,
Lancaster University, UK}} \\
Department of Physics and Astronomy,
University of Manchester, UK \\
and \\
{\bf R.~Kvatadze} \\
High Energy Physics Institute,
Tbilisi State University, Georgia.
\end{center}
\vspace*{2.0cm}

\begin{abstract}
An event weighting method for simulating \BE effects 
in hadronic final states is presented. 
The weight for an event depends on the 
momentum distribution of identical bosons in the event. 
By using a theoretically motivated parametrisation allowing
weights below as well as above unity, the
necessity of a weight-rescaling procedure is eliminated.
A single parameter  
is used to adjust the average event weight to unity. 
Once adjusted, the same 
value of the parameter gives average event weights
that are essentially  
independent of energy, initial quark flavour, 
multiplicity and jet topology. 
The influence of \BE\ correlations on various measurable quantities 
in W pair production is found to be small. In particular, none of the
scenarios considered resulted in a W mass shift larger than
20  MeV.
\end{abstract}

\end{titlepage}

\section{Introduction}
 
There are two main reasons for 
the renewed interest in \BE\ correlations (BEC) in particle 
 physics. One is the  quark-gluon plasma search
 in high energy heavy ion collisions, where BEC are providing important 
 information about the space-time development of the final hadron formation
 in the dense matter. The other is connected to the precision 
 measurement of the W mass in \ee\ annihilation, which can be 
 used to constrain the allowed range of the Higgs boson mass in the Standard
 Model, or restrict the parameter space of any
 other ``new physics''. However, it was suggested that in the fully
 hadronic channel,
 \ee\ $\to$ \WW\ $\to$ \qqqq, BEC and 
 colour reconnection effects could lead to significant uncertainties
 in the determination of W mass, up to ${\cal O}(100$~MeV) 
{\cite{ellis,lonnb}},
 which can effectively render this channel useless and significantly
 reduce the precision on the W mass achievable at LEP2.   

Existing Monte Carlo simulation programs for hadronic final states are 
based on the factorisation property of the QCD amplitudes {\cite{sterm}}: 
the cross
sections are defined  by the perturbative 
parton level amplitudes, while the hadronisation 
process of the final quark states
is simulated  in the framework of a particular model, assuming that
it does not change the probability of the perturbative part.
Various parameters of the hadronisation models have been finely tuned 
to reproduce many aspects of the data, with a notable exception of \BE\
correlations, which cannot be simulated in this approach in principle.

Several attempts have been made to implement 
 Bose-Einstein effects {\it a posteriori}, 
 so that the characteristic BEC are reproduced without breaking down
 the good description of other aspects of the data.
 At the moment, the most popular approach is the one developed 
 in {\cite{lonnb}} and implemented in the PYTHIA Monte 
 Carlo generator {\cite{sjost}}. This model is based on the assumption that 
 the Bose-Einstein effects are local in phase space and are introduced as 
 shifts in final-state boson momenta (hence the rather misleading 
 name of ``local reweighting''). The advantage of this method is 
 that the QCD factorisation is explicitly
 preserved, and, thus, 
 cross-sections of the processes are not affected. This procedure, however, 
 does not conserve energy. In an early approach, energy conservation was 
 restored by rescaling all final-state hadron momenta, while in 
 later algorithms the energies were corrected locally. 
 In this model, a W mass shift arises as a consequence of the final-state 
 boson momentum re-distribution. Numerical predictions depend on the 
 details of the rescaling procedure and vary from 0 up to 100~MeV.
 The width of the W boson is also increased by up to 40~MeV.
 A major disadvantage of this method is that particle momentum 
 re-distribution necessitates the re-tuning of
 the hadronisation parameters in order to reproduce the data, which
 makes the BEC effects rather difficult to extract. 

Another approach to the implementation of Bose-Einstein effects is the
 event weighting method (referred to as ``global reweighting'' in
 \cite{lonnb}). Here, BEC are introduced by assigning weights 
 to the events
 according to the momentum distributions of identical hadrons in the 
 final state.  
Within certain simplifying assumptions,
this procedure can be justified 
 using the formalism  of  Wigner functions {\cite{biala}}.
 A number of such algorithms have been used recently
 to study the 
 Bose-Einstein  effects in the reaction \ee\ $\to$ \WW\ 
{\cite{jadac,kartv,fialk,todor,hakki}} 
(see also {\cite{lonnb, fial1}} for the comparative analysis of different
 methods). 
 Although the various methods differ significantly in 
 the prescriptions for weight calculation, one thing in common to
 all of them is a relatively small BEC-induced shift in the W mass,
 less than about 20~MeV.
 Apart from being much more appealing theoretically,  
 event weighting 
 has also another advantage compared to ``local reweighting'':
 it can, in principle, be applied to the existing Monte Carlo samples.
 
However, the event weighting methods 
 also have some serious shortcomings. The distribution of event weights
 is usually very broad (if not divergent); average weights,
 if taken literally, are usually
 much larger than unity, and in order to keep the cross sections intact 
 a rather arbitrary procedure of weight rescaling is used.
 Average weights may also  vary for various event classes
 such as different initial quark flavours, number of jets in the
 event, multiplicity etc.
 Thus, factorisation is not guaranteed, and is usually preserved 
 by applying  an 
 {\it ad hoc} weight rescaling procedure for 
 each class of events separately. 

These difficulties can be traced back to the fact that event
 weights were larger than unity by construction, implying that BEC enhance 
 configurations where
 identical bosons are close to each other in the phase space. However,
 Bose symmetry can generate repellent forces too, 
 which may become dominant in some areas of the phase space 
 (e. g. identical pions may not exist in a P-wave, so the decay
 $\rho\to\pi^0\pi^0$ is forbidden),
 and give rise to event weights below unity.
 We use a theoretically motivated parametrisation which allows
 some event weights to fall below unity, and thus avoid the necessity
 of weight rescaling. 
 In our method
 the average event weight in \ee annihilation events is adjusted 
 to unity using a single
 parameter, which appears to be independent of energy, initial quark flavour,
 number of jets or particle multiplicity in the event. Inclusive
 spectra of various hadrons also remain unaffected by
 the weighting procedure.   

 The theoretical motivation  
 and description of our method is presented in the following 
 section. 
 The choice of the model parameters is discussed in Section 3.
 In Section 4 the method is applied to \ee 
 annihilation into hadrons in the energy range from 30 to 200~GeV,
 with the region around the Z peak studied in detail.
The influence of 
 Bose-Einstein effects on the process of W pair production is 
 analysed in Section 5. Some conclusion are drawn in Section 6.

\section{Motivation and algorithm description}

Let $M$ be the matrix element 
describing the production of a 
hadronic final state which, among other, non-identical particles,
 contains $n$ identical
bosons. This amplitude consists of $n!$ terms, each corresponding to
a particular permutation $P$ of the $n$ identical particles in the final state:
\beq
M=\sum_P M_P\;.
\eeq
When this process is simulated, the probabilistic treatment of the
hadronisation stage means that the interference between different
amplitudes is not included in the simulation:
\beq
|M|^2_{MC}=\sum_P |M_P|^2 \neq |M|^2.
\eeq
As shown in \cite{bo2}, in order to take
interference terms into account and thus restore the correct
symmetry properties
of the process, a weight $w_P$ has to be assigned to each event:
\beq
|M|^2=\sum_P w_P |M_P|^2,
\eeq
where
\bea{\label{wes1}}
w_P &=& \sum_{P'} {{2{\mathrm{Re}}(M_P M^*_{P'})}\over{|M_P|^2 +|M_{P'}|^2}}
   \nonumber \\
  &=& 1 + \sum_{P'\neq P} {{2{\mathrm{Re}} 
  (M_P M^*_{P'})}\over{|M_P|^2 +|M_{P'}|^2}}\;.
\eea
The sum contains $n!$ terms and depends on the kinematical properties
of the event. However, 
in order to be useful, the above formula needs to be implemented in a recipe
for weight calculation. 

Consider a simple parametrisation for the matrix 
element $M_P$, based on the Lund model of string hadronisation 
\cite{bo2,bo1}:
\beq{\label{pisa}}
M_P=\exp [(i\kappa - b/2) A_P]\;.
\eeq
Here $A_P$ stands for the integral over the space-time area of the
string fragmentation, while $\kappa$ and $b$ are constants 
describing string tension and its breaking probability,
respectively. Substituting (\ref{pisa}) into (\ref{wes1}), 
one obtains:
\beq{\label{wes2}}
w_P = 1 +\sum_{P'\neq P} {{\cos (\kappa \Delta A_{PP'})}\over
                          {\cosh({{b}\over{2}}\Delta A_{PP'})}}\;,
\eeq
where $\Delta A_{PP'} = A_P - A_{P'}$.

In \cite{bo2} it was argued that the dimensionless combination 
$\kappa \Delta A_{PP'}$
between the two configurations
labelled $P$ and $P'$
can be estimated as the scalar product of the 
differences in 4-momentum 
and in the space-time. 
The event weights were calculated at the
stage of event generation by the JETSET program, taking  
the transverse motion of hadrons into account. The resulting 
weights were found to be well-behaved and  described several
manifestations of Bose-Einstein correlations in two-jet events,
but the calculation process is rather labourous and is not easy to
generalise to include more complex jet topologies.

We propose a significantly simplified method of calculating event
weights according to eq. (\ref{wes2}), which, in principle, can be
applied {\it a posteriori} to pre-generated event samples.
We suggest that the combination 
$\kappa \Delta A_{PP'}$
can be estimated as the product of an average interaction radius $R$
and the ``relative momentum'' $Q$, which characterises the difference in
kinematics between the two permutations $P$ and $P'$. 
If the configuration $P'$ is obtained from 
the configuration $P$ by permuting $n$ identical bosons with
masses $m$ and momenta $p_1,\dots,p_n$, then $Q^2$ is
defined as
\beq
Q^2=(p_1 + \dots + p_n)^2 - n^2 m^2,
\eeq
which coincides with the usual definitions $Q^2_{12}=-(p_1-p_2)^2$ and
 $Q^2_{123}=-(p_1-p_2)^2-(p_1-p_3)^2-(p_2-p_3)^2$ for $n=2$ and $n=3$,
respectively.
So, we propose the folowing replacement:
\bea
\kappa \Delta A_{PP'} &\to & RQ, \nonumber \\
{{b}\over{2}} \Delta A_{PP'} &\to & \xi RQ,
\eea
where \x\ is a parameter whose value is to be determined phenomenologically.
This leads to the weight calculated as
\beq{\label{wes9}}
w_P = 1 +\sum_{P'\neq P} {{\cos (RQ)}\over
                          {\cosh(\xi RQ)}}\;.
\eeq
For example, in the simplest case of two identical particles, the weight is
\bea{\label{wes3}}
w_2 = 1 + {{\cos (RQ_{12})}\over{\cosh(\xi RQ_{12})}}\;.
\eea
As noticed in \cite{kartv}, for \x\ values around 1 
the weight (\ref{wes3}), shown
as the solid line in Figure 1, is fairly close to the
Gaussian-type function
\bea{\label{wes4}}
w_G = 1 + \exp(-R^2 Q^2_{12})
\eea
(dashed line in Figure 1), used as the basic weight in 
a number of previous studies {\cite{jadac,kartv,fialk}}. 
The important difference is
that the new basic weight
(\ref{wes3}) goes slightly below unity for some intermediate values
of $Q$,
while the Gaussian weight (\ref{wes4}) always lies above unity.
When the weight of an average event is built
as a product of many terms of the type   
(\ref{wes3}) or (\ref{wes4}), the latter may result in very large
event weights, while the former tends to yield event weights close
to unity.

\begin{figure}\label{fig_curve}
  \begin{center}
    \begin{picture}(100,100)(0,0)
    \epsfig{file=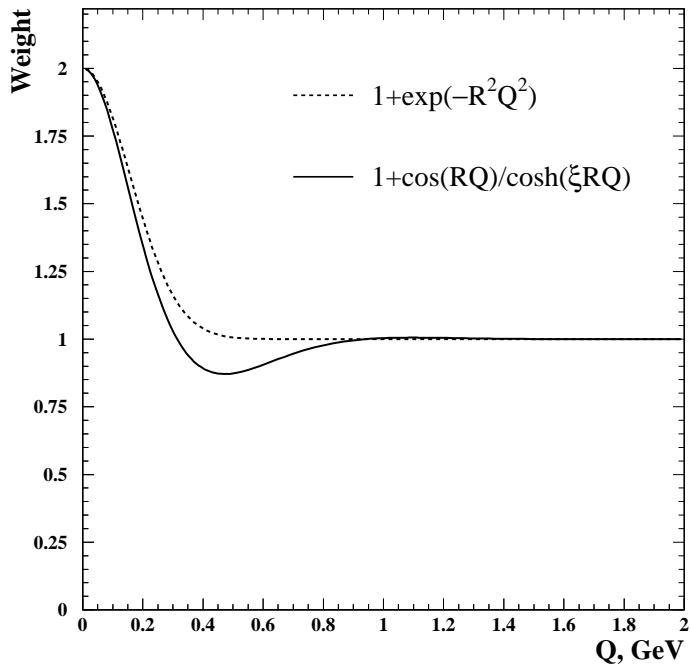,height=100mm}
    \end{picture}
  \end{center}
\caption{\small
The basic weight (\ref{wes3}) used in this paper, compared to the
Gaussian basic weight  (\ref{wes4}), for $R=0.9$~fm and $\xi=1.125$.
}
\end{figure}

Consider, for example, the reaction \ee$\to$\WW, with both Ws decaying
hadronically, and three $\pi^+$ mesons in the final state.  Let the pions
1 and 2 come from the W$^+$ decay while the pion 3 comes
from the W$^-$. The weight for such an event has the following
structure:
\bea{\label{wes5}}
w_3 &=& 1 + (12) + (13) + (23) + 2\times (123)\;,
\eea
where each term stands for one of the six possible permutations.
It has the form $\cos (RQ_{(\alpha)})\;/$ $\cosh (\xi RQ_{(\alpha)})$,
where the numbers in brackets, $(\alpha)=(12),(13),\dots$
 show which pions have been permuted.
The unity corresponds to no permutation, i.e. initial configuration $P$.
The second term describes the only permutation if no inter-W
correlations are allowed, while the following two terms stand for
the two 2-particle inter-W permutations. The last term describes two
3-particle permutations (corresponding to the so called ``genuine''
3-boson correlations \cite{3boson}).

In a final state with 4 identical pions (1 and 2 from W$^+$, 3 and 4 from 
W$^-$) there are 4! terms:
\bea{\label{wes6}}
w_4 &=& 1 + (12) + (34) + (12)(34) + \nonumber \\
    & & (13) + (14) + (23) + (24) +(13)(24)+(14)(23) \\
    & & 2\times[(123)+(124)+(234)+(134)]+6\times(1234)\;. \nonumber 
\eea
The first line contains only permutations of pions originating from 
the same Ws, 
while the remaining terms are either inter-W, or mixed. 
Note a new term type, e.g. (12)(34), corresponding to the 
simultaneous permutation within two pairs of pions. 

In the hadronic final states produced in high energy collisions,
number of identical mesons of each type can be rather big
(e.g. 20 or more in hadronic WW events). 
Of course, only those mesons should be allowed to participate in
BEC, which were produced directly during or shortly after the 
hadronisation phase. 
In the weight calculation we have included only those mesons
whose parents have travelled less than $d_{\mathrm{max}}$ 
in the centre of mass frame.
This excludes mesons from the decays of long-lived
parents, such as $B$ and $D$ mesons, $\tau$ lepton etc.,
leaving on average about 40\% of all mesons.
The remaining mesons should be subject to BEC, but a
straightforward application of the procedure described above would
still lead to serious computational difficulties due to the big number of
permutations. 
However, most of these permuted configurations would have a near-zero
contribution to the weight, as their respective values of the
``distance'' in the momentum space, $Q$, tend to be high. The weight
of each event is essentially determined by clusters of bosons 
with small values of $Q$. 

\begin{figure}\label{fig_dy}
  \begin{center}
    \begin{picture}(100,100)(0,0)
    \epsfig{file=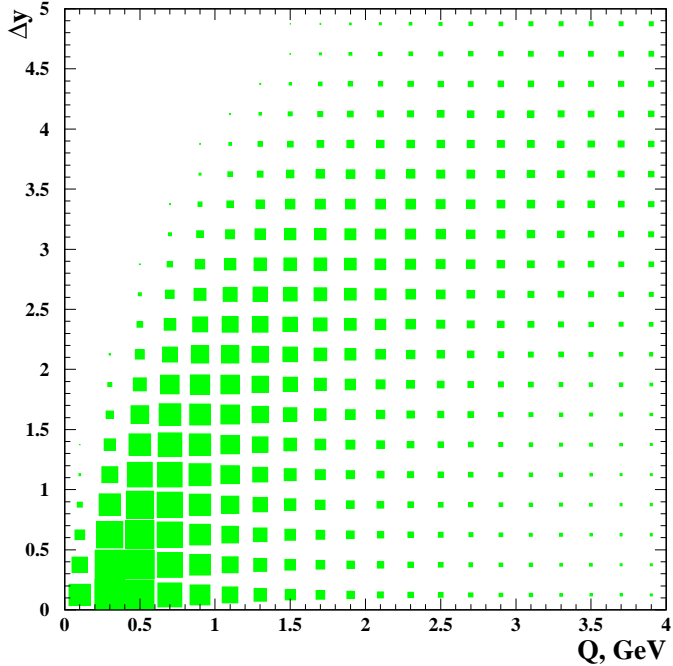,height=100mm}
    \end{picture}
  \end{center}
\caption{\small
Maximum rapidity difference in a cluster of identical charged pions
vs. the variable $Q$ characterising the cluster.
}
\end{figure}

In order to eliminate unnecessary calculations,
we have ordered 
all participating mesons of a particular type
according to their rapidity 
$y=(1/2)\ln[(E+p_z)/(E-p_z)]$ 
(calculated against the thrust axis of the event),
and used the strong correlation existing between the value of
$Q$ characterising the cluster and the 
maximum rapidity difference $\Delta y$
between the mesons in the cluster (see Figure~{\ref{fig_dy}}).
A new cluster was started if the rapidity difference between a
meson and the first meson of the current cluster exceeded
$\Delta y_{\mathrm{max}}$. However, no cluster was allowed to contain
more than $n_{\mathrm{max}}$ mesons. The total weight
for a system of mesons of a particular type was calculated as the product of 
the cluster weights.

Separate weights corresponding to 9 types of mesons ($\pi^+,
\pi^0, \pi^-,K^+,K^0,\overline{K}^0, K^-,\eta, \eta'$) were calculated
for each event, and the event as a whole was assigned a weight 
equal to the product of these 9 weights.

\section{Choice of parameters}

In order to apply the algorithm described above to simulate
BEC in Monte Carlo generated events, the values for the parameters
$R, d_{\mathrm{max}},  n_{\mathrm{max}}, \Delta y_{\mathrm{max}}$ and \x\
have to be fixed. $R$ essentially describes the effective radius of BEC, 
which has been measured in various studies at LEP 
\cite{opalBE,alephBE,delphiBE} 
to be within 0.5 and 1.0 fm. In the following, unless stated otherwise, we
use $R=0.9$~fm. 

The distance $ d_{\mathrm{max}}$ should be small enough 
to exclude from BEC the decay products
of long-lived resonances.
In our studies we used $ d_{\mathrm{max}}=10$~fm.
Note that in our approach the usual parameter $\lambda$, which governs
the ``strength'' of BEC, is missing altogether: identical bosons either
fully participate in BE correlations (if they are produced early during
hadronisation), or do not participate at all (if they come from
 long-lived parents). So the choice of  $ d_{\mathrm{max}}$, which
determines the fraction of participating bosons,
also determines the ``effective strength''  
$\lambda_{\mathrm{eff}}$,
measured by experiments.

The maximum cluster size $n_{\mathrm{max}}$ should be chosen large enough
to allow 2- and 3-particle correlations measured by experiments, but
small enough to keep calculations manageable. 
It was found that the choice $n_{\mathrm{max}}=4$ gives the best overall 
results, and we have used this value in our calculations.
The maximum allowed rapidity difference in a cluster,
$\Delta y_{\mathrm{max}}$, was chosen to be equal to $6/n$, where $n$ is
the number of identical bosons of a particular type in an event. 
This means  $\Delta y_{\mathrm{max}}\lesssim 1 - 1.5$,
which is fairly harmless, as the clusters with larger  $\Delta y$
typically have rather large $Q$ (see Figure~\ref{fig_dy}) and
 their contribution to the event weight is small. 

The parameter $\xi$, which is defined by the ratio of the two scales,
$\kappa$ and $b$, characterising the hadronisation process, determines 
the value of the argument $RQ$ for which the basic weight becomes smaller
than unity. In typical \ee\ events the $Q$-distribution of identical meson
pairs subject to BEC (see Figure~{\ref{fig_Qdis}) 
is such that one can find a value of $\xi$ for which the average event weight
equals unity. In practice, finding this value of \x\ may involve some 
trial-and-error and interpolation, and is only possible up to
a certain precision, determined by Monte Carlo statistics and the variance
of the weight distribution. However, once found, it appears to be
fairly stable under variation of other parameters. 

\begin{figure}\label{fig_Qdis}
  \begin{center}
    \begin{picture}(100,100)(0,0)
    \epsfig{file=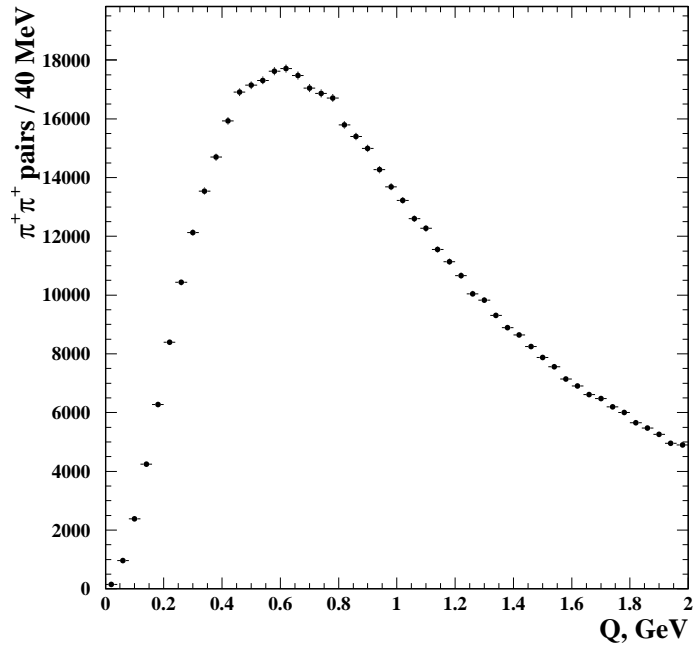,height=100mm}
    \end{picture}
  \end{center}
\caption{\small
$Q$-distribution of identical charged pions subject to BEC in Z decays,
generated with PYTHIA Monte Carlo (see Section 4).
}
\end{figure}


\section{Influence of event weighting on Z properties}

As long as the average event weight for Z hadronic decays  
 is equal to unity, the cross section 
 of this process is not changed by event weighting. However, 
 other measurable properties 
 of the Z could be affected.  

Since the parameters of the Monte Carlo hadronisation models,
 which do not explicitly include BEC effects,
 have been carefully tuned to reproduce various measured
 distributions, uncritical 
 application of event weights may lead to large inconsistencies with 
 measured partonic branching ratios, different jet topologies,
 final hadron multiplicities etc. \cite{kartv}. 
 Also, the average event weight adjusted to
 unity at one energy may deviate from unity at other energies,
 thus potentially affecting such parameters as the mass and the width of 
 the Z boson.

 In order to study how serious these effects are and to judge what 
 consequences they have for the analysis of the WW events, 
 we have compared
 various weighted and unweighted  distributions. 
We have used PYTHIA 6.125 Monte Carlo \cite{sjost} to generate
a sample of $10^5$ hadronic events at  
 $\sqrt{s}=M_{\mbox{Z}}=91.2$~GeV, with \x\ adjusted
 so that the average event weight is equal to 1 (within statistical
 errors).
 The distribution of event weights is shown in Figure~\ref{fig_zwes}.
 It peaks close to its average value and is fairly narrow, with an
 rms of about 0.5. 

\begin{figure}\label{fig_zwes}
  \begin{center}
    \begin{picture}(100,100)(0,0)
    \epsfig{file=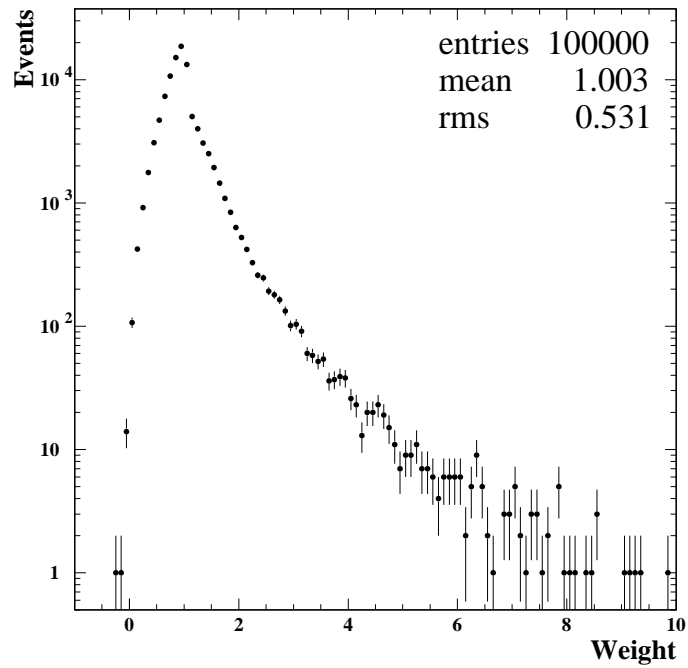,height=100mm}
    \end{picture}
  \end{center}
\caption{\small
Distribution of event weights in Z decays, for $R=0.9$~fm and $\xi=1.125$.
}
\end{figure}
 Table~1 presents average event weights and respective rms values
 for various initial quark flavours and different jet  topologies
 (as determined by the PYCLUS jet finding algorithm with default parameters
\cite{sjost}).
As seen from the table, the average weights are essentially 
 independent of the initial flavour and number of jets in the event.
 This means that the
 implementation of event weights in the form (\ref{wes9}) 
 does not cause any noticeable changes in either the partonic 
 branching ratios or the jet activity in Z decays. 
 This is not trivial, as
 the patterns of the heavy and light quark fragmentation are rather 
 different, and the multiplicity of low-momentum particles is strongly
 correlated with the number of jets in the event. 

Figure~5 shows a very good agreement between the charged particle 
multiplicity distributions at the Z peak 
with (data points with errors) and without (solid line)
event weighting.
Only the errors specific to the weighting process are shown on the plot.
The means of the two distributions differ by  
 $\Delta n_{ch} =0.07 \pm 0.003$, well 
 within the combined experimental error obtained by four LEP 
 experiments, $\pm 0.11$ {\cite{pdp}}.
 Similarly, a very good agreement between
 the weighted (data points)
 and unweighted (solid line)   
 momentum distributions of $\pi^+$ mesons at the Z peak is shown in Figure 6.
 The same is true for other hadron types. 

\begin{figure}\label{fig_mult}
  \begin{center}
    \begin{picture}(100,100)(0,0)
    \epsfig{file=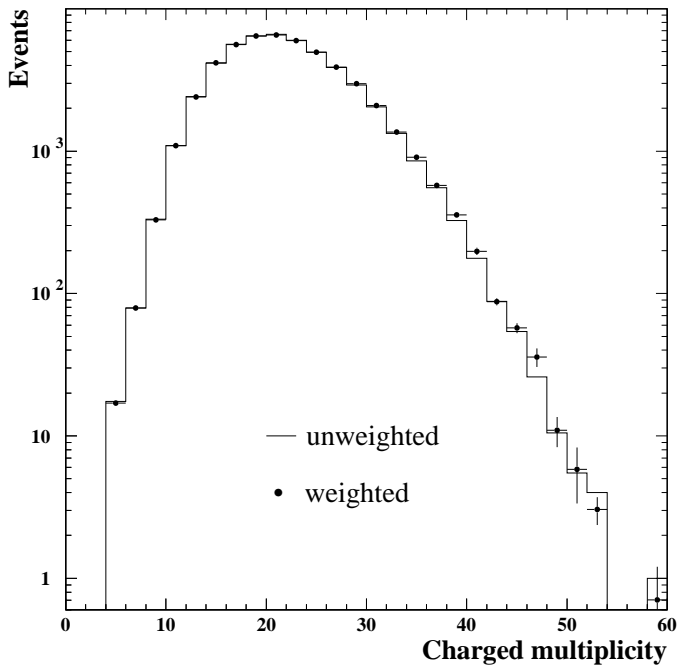,height=100mm}
    \end{picture}
  \end{center}
\caption{\small
The charged multiplicity distribution for events weighted according to our
recipe (data points) compared to the unweighted distribution (solid line). 
The error bars shown correspond 
to the errors specific to the event weighting process, 
and do not include the statistical uncertainties common to both distributions. 
}
\end{figure}

\begin{figure}\label{fig_momu}
  \begin{center}
    \begin{picture}(100,100)(0,0)
    \epsfig{file=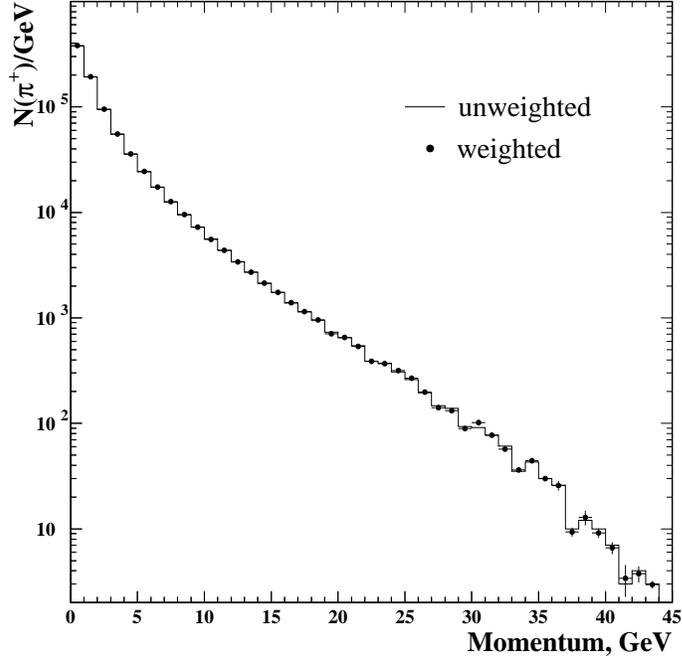,height=100mm}
    \end{picture}
  \end{center}
\caption{\small
The momentum distribution of $\pi^+$ mesons, with events weighted 
according to our
recipe (data points) compared to the unweighted distribution (solid line). 
As in Fig.~5, the error bars 
shown correspond to the errors specific to the event weighting process, and 
do not include the statistical uncertainties common to both distributions. 
}
\end{figure}

 In order to study the dependence
 of the average weight upon initial energy, four more samples of events
 \ee $\to \gamma^*/{\mathrm{Z}}^* \to {\mathrm{hadrons}}$ of
 the same
 size were generated at   
 30, 131, 161 and 200~GeV, using the same value of \x. 
 Average event weights for these energies are also presented in Table~1.
The weights are fairly 
 independent on the initial energy of the collision, 
 although a decrease of about 1\%  
 is seen at the highest energy, 200~GeV.
 This could be connected to the fact that at higher energies a slightly 
 larger percentage of hadronic resonances (such as $\rho$ and $K^*$)
 escape the 10~fm limit, which
 means that their decay products no longer contribute to the weight.

\begin{table}
\begin{center}
\begin{tabular}[tbc]{|c|ccccc|} 
\hline
$q \bar q$ & $d \bar d$ & $u \bar u$ & $s \bar s$ & $c \bar c$ & $b \bar b$ \\
\hline

$\langle w \rangle$ & $1.016 \pm 0.004$ & $1.010 \pm 0.005$ 
 & $1.000 \pm 0.004$ & $1.002 \pm 0.004$ & $0.991 \pm 0.003$ \\
\hline

rms & 0.573 & 0.595 & 0.520 & 0.482 & 0.463 \\
\hline
\hline

$N_{\mathrm{jet}}$ & 2 & 3 & 4 & 5 & 6 \\
\hline

$\langle w \rangle$ & $1.009 \pm 0.003$ & $0.997 \pm 0.002$ 
 & $1.000 \pm 0.003$ & $1.015 \pm 0.007$ & $1.068 \pm 0.023$ \\
\hline

rms & 0.482 & 0.429 & 0.559 & 0.713 & 1.053 \\

\hline
\hline
$E_{\mathrm{cm}}$ & 30~GeV & 91.2~GeV & 131~GeV & 161~GeV & 200~GeV \\
\hline
$\langle w \rangle$ & $0.995 \pm 0.001$ & $1.003 \pm 0.002$ 
 & $0.994 \pm 0.002$ & $0.991 \pm 0.002$ & $0.988 \pm 0.002$ \\
\hline

rms & 0.343 & 0.531 & 0.526 & 0.503 & 0.494 \\
\hline
\end{tabular}
\end{center}
\caption{\small
 Average weights, $\langle w \rangle$, and rms values 
 for different quark flavours, 
 jet topologies
 and centre of mass energies.
}
\end{table}

In order to study possible effects of the event weighting upon the
 mass and the width of the Z resonance, four more samples of
 $10^5$ events were generated at 
 $\sqrt{s}=M_{\mbox{Z}} \pm 2$~GeV and 
 $M_{\mbox{Z}} \pm 4$~GeV, 
 with the same value of the parameter $\xi$. 
 A Breit-Wigner fit to these five points with and without event weighting 
 yielded no significant shifts in
 the Z mass and width:   
 $\Delta M_{\mbox{Z}}=0.4 \pm 0.5$~MeV and
 $\Delta \Gamma_{\mbox{Z}} =1.7 \pm 1.7$~MeV. 
 The experimental uncertainties on these extremely precisely
 measured quantities are 2.2~MeV and 
 2.6~MeV, respectively {\cite{pdp}. 

The correlation functions $C(Q)$ were constructed  as  ratios of the
$Q$ distributions of particle pairs
 for weighted and unweighted events. The ratios
 for same- and opposite-sign charged pion pairs are shown in 
 Figure~7. A clear
 enhancement is seen for the same-sign pion pairs at small values of
 $Q$, while the distribution for the opposite-sign pairs is flat and 
 close to 1.   
 Also shown is the result of the fit to the
 same-sign pair correlation function of the form
 \begin{equation}\label{cq}
   C(Q)=N(1+\beta Q)(1+\lambda_{\mathrm{eff}}
        \exp (-Q^{2}R^{2}_{\mathrm{eff}})),
 \end{equation}
 which is often used to parametrise the experimentally observed correlation
 function in Z decays. The values obtained for the parameters (for a 
 fit range of 0--1 GeV in $Q$) are presented in Table 2, together with the
 results of a similar fit to the
 input pair weight (the latter is merely a fit of the form (\ref{cq}) to the
 basic weight described by (\ref{wes3})). 
The statistical errors in the fitted $Q$
 distributions were increased by 40\%, in order to account for 
 the bin-to-bin correlations, 
 arising from the fact that each boson in an
 event can contribute to several combinations.

\begin{figure}\label{fig_cpion}
  \begin{center}
    \begin{picture}(100,100)(0,0)
    \epsfig{file=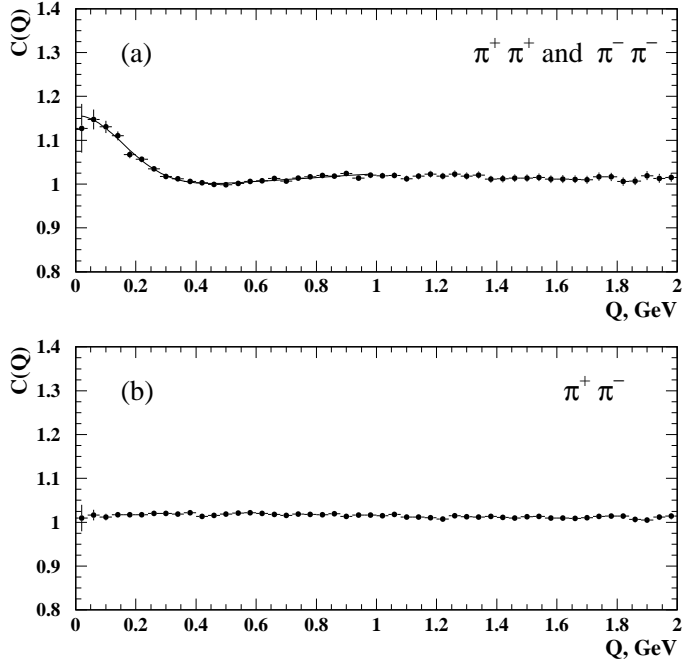,height=100mm}
    \end{picture}
  \end{center}
\caption{\small
The correlation function $C(Q)$, calculated as the ratio of weighted and
unweighted $Q$ distributions, for same-sign charged pions (a) and 
opposite-sign charged pions (b). The line in (a) shows the fit result
using the parametrisation (\ref{cq}), with parameter values given in 
Table 2 (event weight).   
}
\end{figure}

 Fit results for input (pair weight) and output (event weight) 
 parameters are in a reasonable agreement with each other, the only 
 noticeable difference being that the observed enhancement at small $Q$
 is slightly narrower than in the input distribution, leading to a
 larger effective value of $ R_{\mathrm{eff}}$.

So, the event weighting method described above reproduces the
 BE correlation functions 
 without introducing any
 significant modification of the properties of the Z boson,
 and no noticeable energy dependence of the average weight, 
 using the same value for the single adjustable parameter \x\ 
 for all types of events across
 the whole energy range considered.

We have repeated most of our studies for $R=0.6$~fm with very
similar results. The
average event weight was equal to unity (within statistical errors) for
$\xi=0.98$, with no significant dependence on energy, initial quark
flavour or number of jets. The fitted values for the parameters
of the correlation function (\ref{cq}) were:   
$ \lambda_{\mathrm{eff}}=0.29\pm0.03$ and $ R_{\mathrm{eff}}=0.61\pm0.04$.
 The values of these parameters,
 obtained by LEP experiments studying Z hadronic decays,
 vary in the intervals 
 0.2--0.6
 for $\lambda_{\mathrm{eff}}$ 
 and 
 0.5--1.0~fm 
 for $R_{\mathrm{eff}}$, depending on the analysis 
 {\cite{opalBE,alephBE,delphiBE}}.
 More meaningful and detailed comparisons of our results with real data
 will only be possible when the real-world analyses are applied
 to the generator-level Monte Carlo samples, weighted according to our
 recipes. 

\begin{table}
\begin{center}

\begin{tabular}[tbc]{|c|cc|}
\hline
                     & Event weight      & Pair weight \\
\hline
\hline
$\lambda_{\mathrm{eff}}$
            & $0.181\pm 0.012$   & $0.186\pm 0.011$ \\
\hline
$N$             & $0.978\pm 0.008$   & $0.951\pm 0.009$ \\
\hline
$\beta$ (GeV$^{-1}$) & $0.046\pm 0.013$   & $0.053\pm 0.013$ \\
\hline
$R_{\mathrm{eff}}$~(fm) 
                 & $0.902\pm 0.056$   & $0.758\pm 0.038$ \\
\hline
\end{tabular}
\end{center}

\caption{\small
 The fitted values of the output (event weight) and
 input (pair weight)  
 correlation function parameters at the Z peak.
}
\end{table}

\section{BEC in W pair production}

We have studied possible influence of inter-W BE effects 
 on the apparent mass and other measured properties of the W boson.
 The PYTHIA 6.125 event generator \cite{sjost} was again used 
 to simulate the process 
 \ee\ $\to$ \WW\ $\to$ \qqqq, and the weighting method described above 
 was applied to implement Bose-Einstein effects. 
 For obvious reasons, final-state meson distributions in this process 
 are quite different from those in Z decays considered above, and one should 
 expect to obtain the average weight equal to
 unity for a different value of the parameter \x.
 
In weight calculations we have used the same source radius as for Z studies,
 $R=0.9$~fm. Three different weighting schemes were considered:

\begin{itemize}

\item[-]
Only identical bosons originating from different Ws were included in 
 Bose-Einstein correlations (labelled as DW scheme).

\item[-]
Only bosons from the same W are subject to BEC (labelled as SW scheme).

\item[-]
All identical bosons from both the same and different Ws are 
 allowed to participate
 in Bose-Einstein effects (labelled as SW+DW).

\end{itemize}

Two samples of 25000 events were generated at the energies 161 and 200 GeV. 
 We have tuned  $\xi$ to obtain the average value of the event weight
 approximately equal to unity at the energy 161~GeV in each of the 
 three schemes
 separately, and then used the same values of $\xi$ at 200~GeV.
 The application of the parameter $\xi$, fixed at 161~GeV, 
 to the higher energy, 200~GeV, reduces average event weights by about 1.5\%,
 apparently because at higher energies, due to the boost of the W 
 decay frame,  more resonances travel beyond the allowed limit 
 $ d_{\mathrm{max}}$. 
 A similar reduction of the average weight was observed when we applied
 our algorithm to semileptonic WW events, 
  \ee\ $\to$ \WW\ $\to$ \qql\,  at 200 GeV, using the value of 
 \x\ adjusted for Z decays.

Table 3 presents the values of \x,
 average event weights and their rms, 
 together with the shifts in the mass and the width of the W and 
 the average charged particle multiplicity in hadronic W decays,
 compared to the case with no event weighting. The mass and the width
 were determined by fitting 
 a Breit-Wigner parametrisation
 to the invariant mass distribution of the
 W decay products. 

 The fact that the shifts in W parameters 
 for the SW scheme differ from zero shows 
 that our implementation of \BE\ effects is not perfect,
 as we do not know any valid reason
 why the inclusion of BEC only for bosons originating from same W 
 should change any of them 
 (see a similar discussion on Z properties in Section 4). 
 However, we expect that the differences in these quantities 
 between SW+DW and SW represent a valid estimate of the effects
 of inter-W BE correlations, alongside with the predictions of the
 DW scenario. Note that these two sets of shifts (the first and the last 
 rows  for each energy in Table~3) are consistently close to each other.
 Thus, averaging over these two scenarios, our simulations show that 
 the shifts in the W mass, width and
 average charged multiplicity in W decays respectively are
 $13\pm5$~MeV,  $-15\pm 18$~MeV and $0.10\pm0.005$ at 161 GeV,
 reducing correspondingly to 
 $0\pm4$~MeV,  $0\pm 11$~MeV and $0.07\pm0.005$ at 200 GeV.
 Typical experimental errors on these quantities
 are at present significantly larger: 56~MeV, 50~MeV and 0.4 \cite{pdp}.
     
\begin{table}
\begin{center}

\begin{tabular}[tbc]{|c|cccrrc|}
\hline
                 & $\xi$ & $\langle w \rangle$ & rms & $\Delta M_W$
& $\Delta \Gamma_W$ & $\Delta n_{ch}$ \\
                 &  & &  & (MeV) & (MeV) &  \\
\hline
\hline
161~GeV        &  &  &  &  &  & \\  
\hline
\hline
DW               & 1.008 & $0.989 \pm 0.003$ & 0.491 & $13 \pm 5$
& $-11 \pm 18$   & $0.12 \pm 0.005$ \\    
\hline
SW               & 1.048 & $1.007 \pm 0.004$ & 0.679 & $-2 \pm 6$
& $-45 \pm 26$   & $0.19 \pm 0.005$ \\
\hline
SW+DW            & 1.094 & $1.002 \pm 0.006$ & 0.930 & $11 \pm 9$
& $-63 \pm 33$   & $0.27 \pm 0.005$ \\
\hline
(SW+DW)-(SW)     &       &                   &       & $13 \pm 6$
& $-18 \pm 20$   & $0.08 \pm 0.005$ \\
\hline
\hline
200~GeV        &  &  &  &  &  & \\  
\hline
\hline
DW               & 1.008 & $0.983 \pm 0.003$ & 0.430 & $1 \pm 4$
& $3 \pm 11$     & $0.08 \pm 0.005$ \\    
\hline
SW               & 1.048 & $0.986 \pm 0.004$ & 0.623 & $-5 \pm 5$
& $-34 \pm 15$   & $0.11 \pm 0.005$ \\
\hline
SW+DW            & 1.094 & $0.986 \pm 0.005$ & 0.814 & $-7 \pm 6$
& $-37 \pm 20$   & $0.17 \pm 0.005$ \\
\hline
(SW+DW)-(SW)     &       &                   &       & $-2 \pm 4$
& $-3 \pm 13$    & $0.06 \pm 0.005$ \\ 
\hline
\end{tabular}
\end{center}

\caption{\small
 Parameter $\xi$, average weights and their rms for \WW\ hadronic decay events
 at 161 and 200~GeV, together with shifts in W mass, width and 
 charged multiplicity caused by event weighting.
}
\end{table}

Table 4 presents values of the average event weight and 
 the rms of the event weight distribution, 
 for varous jet topologies in the DW scenario.
 The average weights are essentially independent of 
 the number of jets.
 Hence, the implementation of event weights does not change 
 significantly the jet multiplicity distribution in the \WW\ production 
 process. The rms of the weight distribution  
 increases slightly with the number of jets, as in Z decays. 
 This is connected to 
 the increase of particle multiplicity with increasing  number of jets. 
 The event weight dependence on the flavours of the initial quarks in 
 W decays is very weak and does not change partonic branching ratios. 
 We have also checked that the introduction of event weights does not
 alter multiplicity distributions and inclusive momentum spectra for 
 various hadrons in \WW\ production.
 Similar results were obtained also for SW+DW and SW
 event weighting schemes.   

\begin{table}
\begin{center}
\begin{tabular}[tbc]{|c|ccccc|} 
\hline

$N_{\mathrm{jet}}$ & 4 & 5 & 6 & 7 & 8 \\
\hline
\hline
161~GeV      &   &   &   &   &   \\
\hline
\hline
$\langle w \rangle$ & $0.964 \pm 0.008$ & $0.969 \pm 0.005$ 
 & $0.982 \pm 0.005$ & $1.001 \pm 0.007$ & $1.030 \pm 0.012$ \\

rms & 0.352 & 0.410 & 0.443 & 0.558 & 0.641 \\

\hline
\hline
200~GeV      &   &   &   &   &   \\
\hline
\hline
$\langle w \rangle$ & $0.968 \pm 0.009$ & $0.970 \pm 0.005$ 
 & $0.978 \pm 0.005$ & $0.984 \pm 0.005$ & $1.000 \pm 0.008$ \\

rms & 0.319 & 0.344 & 0.393 & 0.430 & 0.483 \\

\hline
\end{tabular}
\end{center}
\caption{\small
 Average event weights and the rms for different number of jets at 161 and
 200~GeV, in the DW scheme.
}
\end{table}

On the experimental side, it is still not clear at the moment 
whether the \BE\ correlations  
 between mesons originating from different Ws exist or not
 {\cite{aleph,opal,l3,delphi}}. Therefore, we have studied the correlation 
 functions for identical 
 bosons in all three scenarios (DW, SW and SW+DW).  
 The correlation functions were constructed as ratios of particle pair 
 four-momentum difference distributions for weighted and unweighted \WW\ 
 hadronic decay events. 
 They are plotted in Figure~8 for the charged pion case.
In all three scanarios, the figure shows a clear enhancement at small
 $Q$, characteristic of \BE\ correlations.

\begin{figure}\label{fig_cpiww}
  \begin{center}
    \begin{picture}(100,100)(0,0)
    \epsfig{file=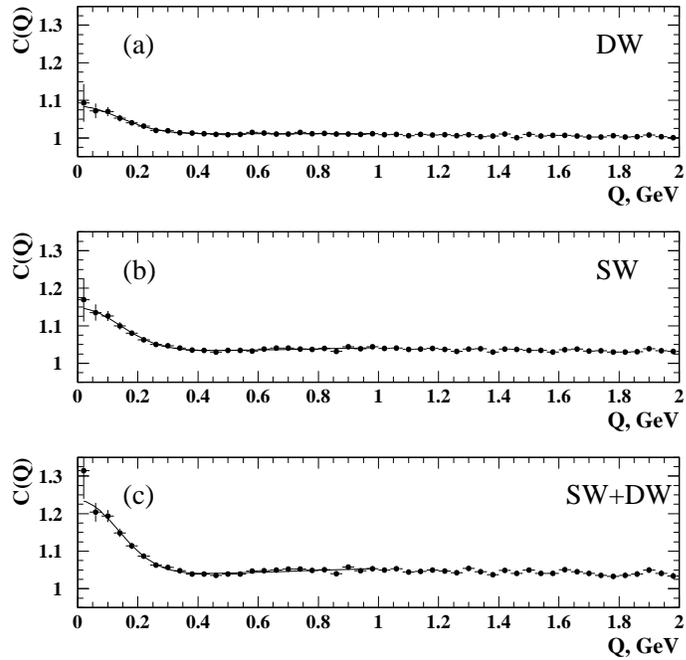,height=100mm}
    \end{picture}
  \end{center}
\caption{\small
The correlation functions $C(Q)$, 
for the three different scenarios described in the text:
a) only bosons from different Ws participate in BEC (DW);
b) only bosons from the same W participate in BEC (SW);
c) all identical bosons participate in BEC (SW+DW).
The lines show the fit results
using the parametrisation (\ref{cq}), with parameter values given in 
Table 5.   
}
\end{figure}

 Values for the effective strength of correlations $\lambda_{\mathrm{eff}}$  
 and the observed effective source radius $R_{\mathrm{eff}}$, obtained 
 by fitting the form (\ref{cq}) to these distributions, are given in Table 5.
The strength of 
 the correlations in the DW case is significantly smaller 
 compared to the full SW+DW 
 scenario,  because a large
 number of identical bosons come from the same W, and the pairs 
 of identical bosons from different Ws are on average farther away from
 each other, in the $Q$ space, than those from the same Ws. 
 Hence, the experimental observation of Bose-Einstein correlations 
 for charged pion pairs originating from different Ws at LEP2 would 
 indeed be very difficult, requiring high statistics, 
 and a careful control 
 of systematics coming from the choice of the reference sample. 
 As in the case of hadronic Z decays, the effective input (pair weight) and 
 output (event weight) values for
 parameters  $\lambda_{\mathrm{eff}}$ and $R_{\mathrm{eff}}$
 are reasonably close to each other.   
\begin{table}
\begin{center}
\begin{tabular}[tbc]{|c|ccc|}
\hline
          & DW              & SW              & SW+DW \\
\hline
\hline
161~GeV &                 &                 & \\
\hline
\hline
$\lambda_{\mathrm{eff}}$ 
       & $0.073\pm .012$ & $0.117\pm .013$ & $0.201\pm .015$ \\
\hline
$R_{\mathrm{eff}}$~(fm)  
       & $1.026\pm .126$ & $1.006\pm .086$ & $1.053\pm .057$ \\
\hline
\hline
200~GeV &                 &                 & \\
\hline
\hline
$\lambda_{\mathrm{eff}}$ 
      & $0.057\pm .010$ & $0.116\pm .012$ & $0.182\pm .014$ \\
\hline
$R_{\mathrm{eff}}$~(fm)  
     & $0.902\pm .107$ & $0.994\pm .081$ & $0.993\pm .055$ \\
\hline
\end{tabular}
\end{center}

\caption{\small
The fitted values of the correlation function parameters 
$\lambda_{\mathrm{eff}}$ and $R_{\mathrm{eff}}$ 
for W pair production in (DW), (SW) and (SW+DW) schemes at 
 $\protect\sqrt{s} = 161$ and 200~GeV.
}
\end{table}

\section{Conclusions}

We have developed a method for modelling Bose-Einstein correlations
 using event weighting, with a theoretically motivated parametrisation
 for the basic weight, based on the string fragmentation picture. 
 Our approach differs from some other 
 ``global weighting''
 schemes, as the event weights in our case are distributed both above and
 below unity 
 with a rather narrow
 distribution,
 and the average event weight is easily adjusted to unity using a single
 parameter \x.  
 This eliminates the additional 
 rescaling of event weights, necessary in models where 
 all event weights are larger than one. 

By weighting Monte Carlo events in accordance with our prescriptions,
 the experimentally observed 
 characteristic \BE\ enhancement at small relative momenta is
 reproduced.
 Good agreement was found between the input and
 output values of the parameters 
$\lambda_{\mathrm{eff}}$ and $R_{\mathrm{eff}}$
 in the de-facto
 standard Gaussian parametrisation. By fine-tuning the values of
 our model parameters $d_{\mathrm{max}}$ and $R$ (and subsequent 
 re-adjustment of \x\ in order to keep the average weight equal to unity)
 one can bring  $\lambda_{\mathrm{eff}}$ and $R_{\mathrm{eff}}$ closer to the
 values obtained by particular experiments. However, any  
 detailed comparison of our results with the
 real data will only be possible when the real-world analyses are applied
 to the  Monte Carlo samples, weighted according to our
 recipes.

The main weakness of all BEC implementations via event reweighting
 is the possible violation of 
 factorization between the hard perturbative part of the process and the
 non-perturbative hadronisation stage,
 essential to all Monte Carlo generators.
 Our model is practically free from these difficulties.
 We have made extensive checks by comparing our predictions
 with unweighted distributions, which have been tuned and tested to
 reproduce very precise experimental data on Z decays. 
 We have found no significant
 shifts in the mass, width and partonic branching fractions 
 of the Z boson due to event reweighting, within the estimated
 errors which are well below the level of existing experimental
 uncertainties. The same is true for charged multiplicity
 distributions and inclusive spectra of various final-state hadrons.

In the process \ee\ $\to$ \WW\ $\to$ \qqqq, the introduction of event weights
 leads to small shifts in the values of W mass, width and charged multiplicity,
 less than 15~MeV, 20~MeV and 0.10, respectively at $\sqrt{s}=161$~GeV, 
 and even smaller at 200~GeV. These values are well below currently
 existing experimental errors.
Hence, at the generator level the  
 Bose-Einstein correlations as implemented in our model 
 do not introduce large additional uncertainties 
 in the determination of W characteristics in the fully hadronic channel. 
 However, as in the case of Z decays, 
 in order to assess the effects of BEC on the experimentally measured W 
 parameters, the detector simulation and the actual fitting procedures 
 used by the LEP experiments
 have to be applied to the weighted event samples. This is not too
 difficult because there is no need to generate special Monte Carlo 
 samples, as the weighting can be applied {\it a posteriori}\ \
 to the existing Monte Carlo events. 

We have also studied the correlation 
 functions for charged pions in fully hadronic WW events
 in three separate scenarios, depending on which pairs of identical 
 bosons were allowed to participate in BEC: only from different Ws  
 (DW), only from the same W (SW) and all pairs (SW+DW).
The characteristic  
 enhancement at $Q\lesssim 0.2$~GeV was seen in all three scenarios.  
 However, the effective value of the parameter $\lambda_{\mathrm{eff}}$
 is the smallest 
 in the (DW) scenario, suggesting that
 the direct observation  of
 BEC for pions originating from different Ws
 with the available statistics from LEP2 would be difficult,
 requiring a very careful control of systematics.

Helpful discussions with N.~Kjaer, G.Lafferty and L.~L\"{o}nnblad 
are thankfully acknowledged.
One of us (R.K.) would like to thank 
 the Particle Physics Group at the
 University of Manchester, and IPPP (Durham) for their kind hospitality.
 
 This work was supported in part by the INTAS grant 99-0558.

\end{document}